\documentclass[twocolumn,amsmath,amssymb,aps,pra,longbibliography,floatfix]{revtex4-1}
\usepackage{physics}
\usepackage{units}
\usepackage{amsmath}
\usepackage{cases}
\usepackage{url}
\usepackage{natbib}
\usepackage{textcase}
\usepackage{amssymb}
\usepackage{graphicx}
\usepackage{bm} 

\usepackage{times}
\usepackage{float}
\usepackage{multirow,microtype,color,relsize}
\usepackage{subfigure}
\usepackage[breaklinks=true]{hyperref}
\usepackage[utf8]{inputenc}
\usepackage[english]{babel}
\hypersetup{colorlinks=true,linkcolor=blue,citecolor=blue}
\hypersetup{linktocpage}
\usepackage{CJK}
\usepackage{breakcites}
\usepackage{float}
\usepackage{siunitx}
\usepackage[dvipsnames]{xcolor}
\definecolor{mypine}{RGB}{1, 121, 111}

\begin{document}
\begin{CJK*}{UTF8}{gbsn}
\title{Spectral rigidity of non-Hermitian symmetric random matrices near Anderson transition}

\author{Yi Huang~(黄奕)}
\email[Corresponding author: ]{huan1756@umn.edu}
\author{B.\,I. Shklovskii} 
\affiliation{School of Physics and Astronomy, University of Minnesota, Minneapolis, Minnesota 55455, USA}

\received{\today}

\begin{abstract}

We study the spectral rigidity of the non-Hermitian analog of the Anderson model suggested by Tzortzakakis, Makris and Economou (TME). 
This is a $L\times L \times L$ tightly bound cubic lattice, where both real and imaginary parts of on-site energies are independent random variables uniformly  distributed between $-W/2$ and $W/2$. The TME model may be used to describe a random laser. In a recent paper we proved that this model has the Anderson transition at $W= W_c \simeq 6$ in three dimension. 
Here we numerically diagonalize TME $L \times L \times L$ cubic lattice matrices and calculate the number variance of eigenvalues in a disk of their complex plane. 
We show that on the metallic side $W < 6$ of the Anderson transition, complex eigenvalues repel each other as strongly as in the complex Ginibre ensemble only in a disk containing $N_c(L,W)$ eigenvalues. We find that $N_c(L,W)$ is proportional to $L$ and grows with decreasing $W$ similarly to the number of energy levels $N_c$ in the  Thouless energy band of the Anderson model. 

\end{abstract}

\date{\today}

\maketitle
\end{CJK*}

\section{Introduction} 

The theory of random matrix spectra was first applied to the nuclear and atomic physics ~\cite{wigner:1951,*wigner:1955,*wigner:1957,*wigner:1958,dyson:1962a,*dyson:1962b,*dyson:1962c,dyson:1963,porter:1965}.
Since then, many applications has been developed in other fields, in particular, in the condensed matter physics~\cite{guhr:1998,beenakker:1997,*beenakker:2011}. 
One of the standard measures of the level statistics is the spectral rigidity, defined as the level number variance $\Sigma^2 = \ev{N^2} - \ev{N}^2$ in an energy 
interval of length $E$ containing on average $\ev{N}$ levels. For a Gaussian orthogonal ensemble (GOE), Dyson~\cite{dyson:1962b} showed that at large $\ev{N}$ the number variance is given by $\Sigma^2 \simeq (2/\pi^2) \log \ev{N}$. 

Later Altshuler and Shklovskii found that the spectral rigidity is related to the fluctuations of conductance of small metallic samples with random impurities~\cite{altshuler:1986}. 
They showed that in a metallic cube with a side $L>>l$, where $l$ is the mean free path, Dyson's result is valid only if $E \leq E_c$, where $E_c= \hbar D/L^2$ is the Thouless energy and $D$ is the diffusion coefficient of electrons in the metal. 
At $E \gg E_c$ they predicted that $\Sigma^2 \propto E^{3/2}$. 
Numerical diagonalizations of the Anderson Hamiltonian~\cite{anderson1958} confirmed the former prediction, but showed that an observation of $\Sigma^2 \propto E^{3/2}$ requires diagonalization of Anderson model for cubes with unrealistically large $L$. 
In the achieved crossover range  $E_c < E < 100 E_c$, it was found that $\Sigma^2 \simeq E/E_c$ ~\cite{braun:1995,zharekeshev:2001}. 
This $\Sigma^2 \simeq E/E_c$ and Dyson's result themselves are much smaller than the one for Poissonian random levels $\Sigma^2 = \ev{N}$, which is valid for the insulating limit of the Anderson Hamiltonian. 
This is of course a result of level repulsion.

Ref.~\cite{altshuler:1988} studied how $\Sigma^2$ evolves with the increasing disorder parameter $W$ of the Anderson model when one crosses metal-insulator transition. 
It was shown that on the metal side at $W < W_c$, this evolution can be described by a decrease of energy $E_c$ following a decrease of $D$.
This is because $D \propto v_F l$ and $l \propto W^{-2}$.
At the metal-insulator transition, $W=W_c$, one gets semi-Poissonian statistics $\Sigma^2 = \kappa \ev{N}$ with $\kappa \simeq 0.25$, which is smaller than  $\kappa=1$ for the Poisson limit of the insulating phase achieved only at $W\gg W_c$~\cite{altshuler:1988,Bogomolny:2001}.

Recently, research interest in the metal-insulator transition has moved to non-Hermitian systems such as random lasers~\cite{wiersma:2008,*wiersma:2013,schonhuber:2016,basiri:2014}, biological networks~\cite{nelson:1998,rajan:2006,amir:2016,zhang:2019}, and spin chains~\cite{oganesyan:2007,hamazaki:2019a,sa:2020}.
For non-Hermitian systems with asymmetric hopping matrix elements between lattice sites, delocalization of wave functions is possible in dimensions less than three. 
Examples are Hatano-Nelson matrices demonstrating delocalization transition even in one dimension~\cite{hatano:1996}. 

Another class of random Non-Hermitian matrices with symmetric hopping was introduced by Tzortzakakis, Makris and Economou (TME)~\cite{tzortzakakis:2020}. 
They suggested a simple and elegant extension to the conventional Anderson model to non-Hermitian matrices. 
It consists of a tight-binding lattice with real symmetric overlap energies $I_{ij} = 1$, and random complex onsite dimensionless energies $E_i$ whose real and imaginary parts are independent random variables distributed uniformly between $-W/2$ and $W/2$.
The Hamiltonian reads
\begin{equation}\label{eq:h}
	H = \sum_i E_i a^{\dagger}_i a_i - \sum_{i,j} a^{\dagger}_i a_j,
\end{equation}
where $i,j$ in the second term are nearest neighbors, and no bonds outside the cube surface. 
TME model may describe a random laser media with balanced in average random local loss and amplification. 
TME studied numerically many realizations of two-dimensional square 50 $\times$ 50 lattices with different values of $W$ and noticed a tendency to delocalization of wave functions with decreasing $W$ from 5 to 1. 

TME work lead us~\cite{huang:2020} to explore whether the TME model has the Anderson metal-insulator transition with growing $W$ in two and three dimensions. 
For this purpose we used the nearest neighbor spacing statistics of complex eigenvalues of TME matrices. 
Namely, we studied the ratio of the first and second nearest neighbor spacing $r_L(W)$ as a function of the disorder strength $W$ and the size of the sample $L^{d}$ where $d$ is the dimensionality. 
We found that at $d =3$ the curves $r_L(W)$ for different $L$ cross at $W = W_c = 6.0 \pm 0.1$, signalling that at $d=3$ the Anderson transition exists and happens at $W = W_c = 6.0 \pm 0.1$. 
However, at $d=2$ there is no such crossing so that at any finite $W$ all eigenstates are localized.
Thus, the Anderson transition exists in the three dimensional TME model, but is missing in two dimensions, similarly to the conventional Anderson model. 
This finding was confirmed by the scaling theory of non-Hermitian localization which emphasized the important role of the reciprocity symmetry~\cite{kawabata:2020}.

In this paper we continue the study of the level statistics of TME matrices and explore the behavior of the spectral rigidity near the Anderson transition. 
To the best of our knowledge, the spectral rigidity of non-Hermitian random matrices was fully explored only for the Ginibre ensemble~\cite{ginibre:1965,jancovici:1981,*jancovici:1993,levesque:2000}~\footnote{The complex Ginibre ensemble is equivalent to a two-dimensional one-component plasma at temperature $T = (\pi\rho)^{1/2}e^2/2k_B$, where $e$ is the unit charge and $\rho$ is the density of the plasma. The proof of this equivalence can be seen, for example, in Chapter 15 of Ref.~\onlinecite{forrester:2010}.}.
Here, we numerically calculate the number variance $\Sigma^2$ in the three dimensional TME model and study the evolution of $\Sigma^2$ with $W$ as the system goes from a metal to an insulator.
We use statistics of complex eigenvalues obtained by diagonalization of the TME model on many realizations of $L \times L \times L$ cubic lattices with $L = 8,12,16$.
The diagonalization is done using LAPACK algorithm~\cite{LAPACK}.
Unlike the real spectrum of Hermitian systems, now the eigenvalues are points in the two dimensional complex plane.
Therefore, we select a disk of radius $E$ centered at the origin of the complex plane, and study how $\Sigma^2$ depends on the average number of eigenvalues $\ev{N}$ inside such a disk.

\section{Number variance}
\begin{figure}[t]
    \centering
    \includegraphics[width = \linewidth]{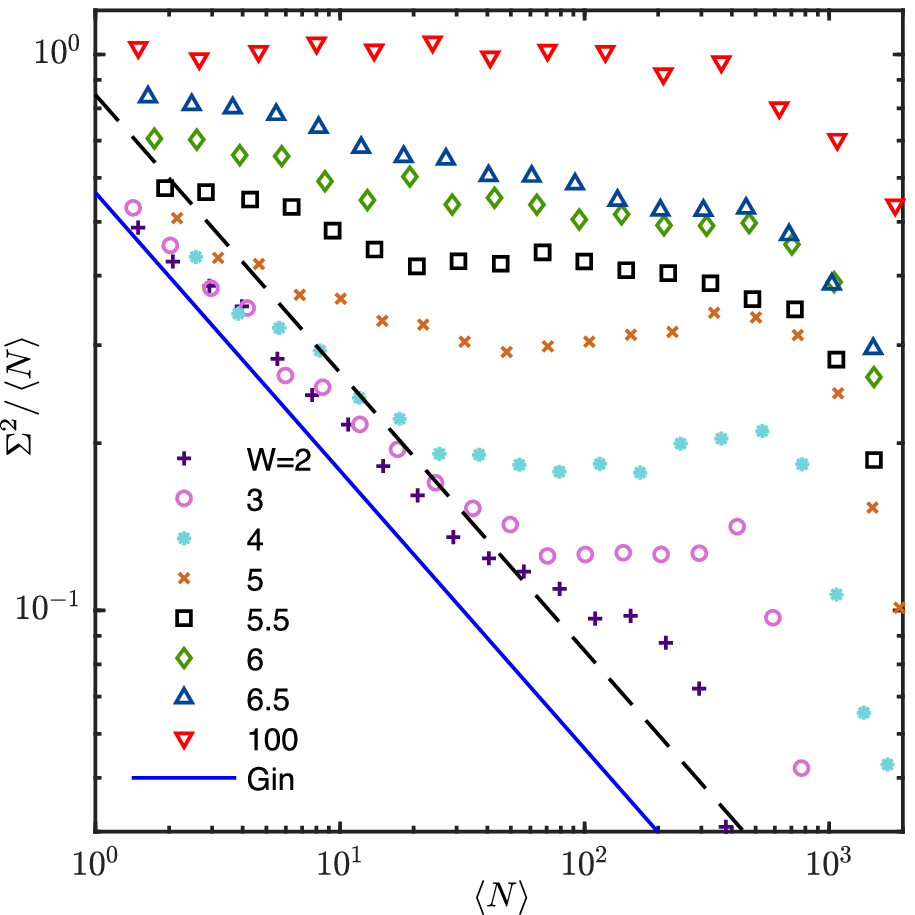}
    \caption{(Color online). $\Sigma^2/\ev{N}$ as a function of $\ev{N}$ for 1000 realizations of a cube $16 \times 16 \times 16$ for different $W$. The blue solid line represents the analytical result for the complex Ginibre ensemble $\Sigma^2/\ev{N} = (\pi\ev{N})^{-1/2}$~\cite{ginibre:1965,jancovici:1981,*jancovici:1993,levesque:2000}. The dashed line shows $\Sigma^2/\ev{N} = 1.5(\pi\ev{N})^{-1/2}$. Numbers of strongly repelling each other eigenvalues $N_c(16,W)$ are defined by crossing points of the dashed line and extrapolated line through the data points for a given $W$.}
    \label{fig:var}
\end{figure}

In Figure~\ref{fig:var} our results for $\Sigma^2/\ev{N}$ are plotted as a function of $\ev{N}$ in log-log scale at $L=16$ and $W=$2, 3, 4, 5, 5.5, 6, 6.5 and 100.
We see that these results qualitatively remind ones for the Anderson model~\cite{braun:1995,zharekeshev:2001}. 
At the transition point $ W= W_c = 6$ the ratio $\Sigma^2/\ev{N} \simeq 0.5$ approximately showing semi-Poissonian statistics. 
At larger $W$ there is a crossover between the semi-Poissonian and Poissonian statistics. 

On the metallic side of the transition $W < 6$ we see that at small  $\ev{N}$ all curves are close to the complex Ginibre ensemble value $\Sigma^2/\ev{N} = (\pi\ev{N})^{-1/2}$~\cite{ginibre:1965,jancovici:1981,*jancovici:1993,levesque:2000}~\footnote{We are not aware of a proof that our number variance at small disorder should be identical to that of the complex Ginibre ensemble. However, within the accuracy of our numerical calculations, they are empirically indistinguishable.}.
Thus our numerical data shows that the Ginibre value plays the role similar to the Dyson limit $\Sigma^2/\ev{N} =2\log \ev{N}/\pi^2\ev{N} $ in the Anderson model~\cite{braun:1995,zharekeshev:2001}.

As in the case in the Anderson model, at very small $ W$ the mean free path $l \sim (W_c/W)^2$ becomes larger than the system size $L$ and the transport becomes ballistic. 
In this case, disorder only acts as small perturbation~\cite{sivan:1987}, the spectrum is determined by the quantization of the tight-binding model in the cube and has nothing to do with disorder induced chaotic motion. Parametrically, this happens only at $W/W_c < L^{-1/2}$. In our case, this inequality is violated even at $W=2$ and indeed we see chaotic results close to ones of the complex Ginibre ensemble.

One can interpret the origin $\Sigma^2 \propto \ev{N}^{1/2}$ in our data and in the complex Ginibre ensemble limit in the same way as the Dyson's result is interpreted in the Ref.~\cite{dyson:1962a,altshuler:1986}. 
Let us think about complex eigenvalues  $\epsilon_i$ as the Coulomb gas of particles interacting via logarithmic repulsion $U(\epsilon_i - \epsilon_j) \propto -\log(\abs{\epsilon_i - \epsilon_j})$. 
This gas is confined by an external potential in the complex plane at a temperature of order of the repulsion at average distance between nearest neighbors. 
Logarithmic interaction allows thermal fluctuation to separate a ``particle'' and its ``vacancy'' only by the average distance between ``particles''. 
Thus, fluctuations of number of particles in the disk happen only due to local independent fluctuations along its border, each having a random sign and the mean square value of the order unity. 
A disk containing on average $\ev{N}$ levels has perimeter proportional to $\sqrt{\ev{N}}$, and a number of such independent contributions should be proportional to $\sqrt{\ev{N}}$, which explains $\Sigma^2 \propto \sqrt{\ev{N}}$ observed in Figure~\ref{fig:var} for the TME model at small enough $\ev{N}$.

\section{Critical number \texorpdfstring{$N_c$}{Nc}}
At larger $\ev{N}$ we see that for all $W$ the ratios $\Sigma^2/\ev{N}$ deviate from the complex Ginibre ensemble limit. 
To characterize this deviation we introduce the critical number $\ev{N} =N_c(L,W)$ where in Figure~\ref{fig:var} the ratio $\Sigma^2/\ev{N}$ becomes larger than its complex Ginibre ensemble limit by 50\%.
To help to find $N_c(L,W)$ we added the dashed line $\Sigma^2/\ev{N} = 1.5(\pi\ev{N})^{-1/2}$ in Figure~\ref{fig:var}. 
Thus, $N_c(L,W)$ are defined by crossing points of the dashed line and extrapolated lines through the data points for a given $W$.

\begin{figure}[t]
    \centering
    \includegraphics[width = \linewidth]{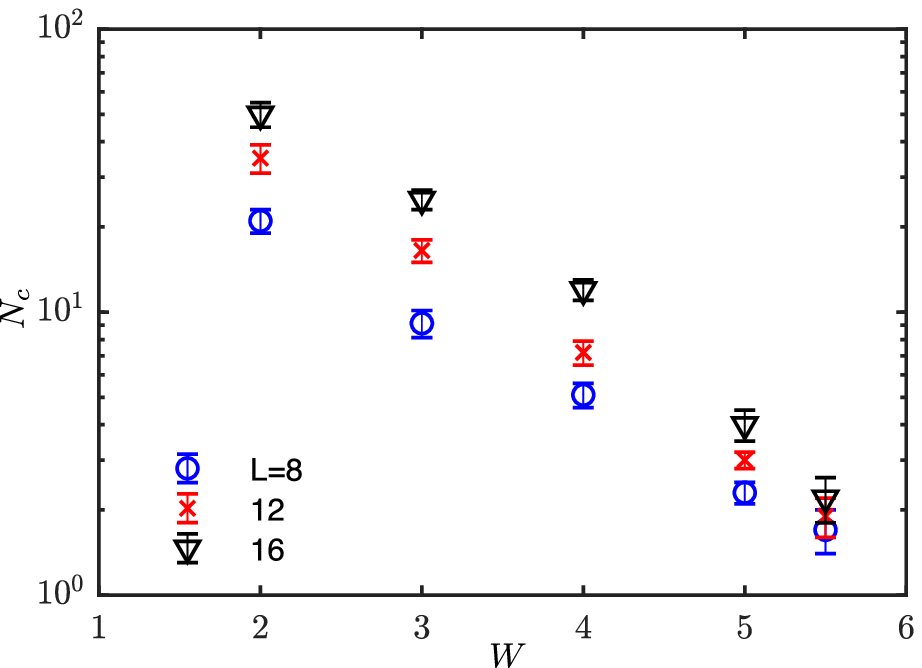}
    \caption{(Color online). The characteristic number $N_c$ as a function of $W$ for $L = 8, 12, 16$.} 
    \label{fig:nc1}
\end{figure}

\begin{figure}[t]
    \centering
    \includegraphics[width = \linewidth]{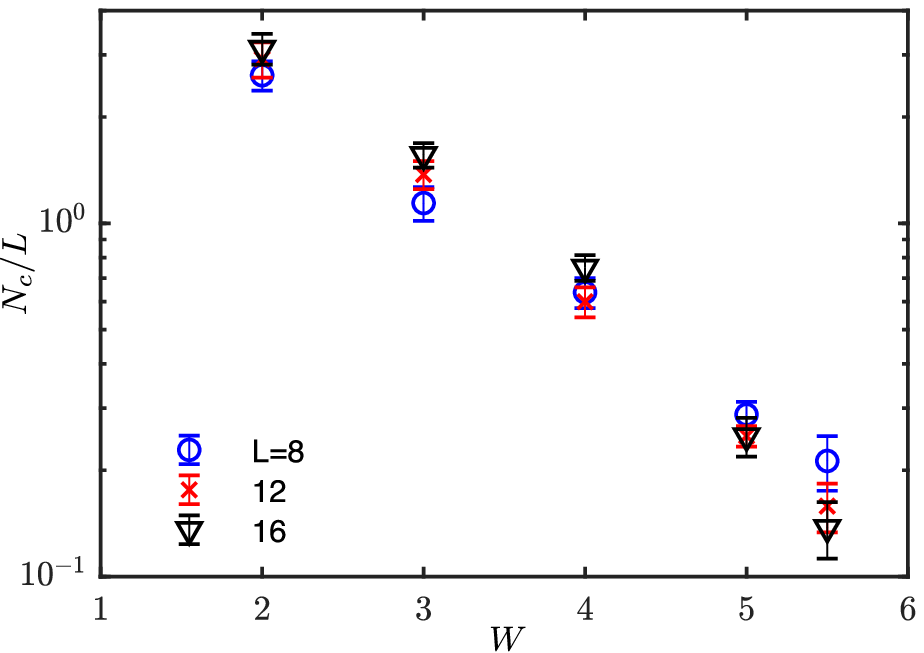}
    \caption{(Color online). The ratio $N_c/L$ as a function of $W$ for $L = 8, 12, 16$.} 
    \label{fig:nc}
\end{figure}

Our results for $N_c(L,W)$ are shown in Figure~\ref{fig:nc1} for $L=16$ and two smaller cube with size $L=8$ and $L=12$. 
To provide better statistics in the latter cases we used 40000 and 12000 realizations respectively. 
We see that all $N_c(L,W)$ are close to unity at the transition point $W=W_c$ and grow with decreasing $W$ when samples become more metallic. 
We also see strong dependence $N_c(L,W)$ on $L$. Figure~\ref{fig:nc} shows the ratios $N_c(L,W)/L$ for the metal side of the transition $W < W_c=6$. 
The same values of $N_c(L,W)/L$ for different $L$ at $W \leq 5$ clearly show the scaling $N_c(L,W) \propto L$. 
At $W=5.5$ we are already in the critical vicinity of the Anderson localization transition and the scaling $N_c(L,W) \propto L$ fails. 

Figure~\ref{fig:var} shows that the ratio $\Sigma^2/\ev{N}$ after the departure from the complex Ginibre ensemble limit has a long plateau at $\ev{N} > N_c$.
In other words, this means that there is a crossover range of  $\ev{N} > N_c$ where $\Sigma^2 \propto \ev{N}$. 
This behavior reminds the mentioned above linear growth $\Sigma^2 \simeq E/E_c$ for the Hermitian case~\cite{braun:1995,zharekeshev:2001}. 
The sharp drop of $\Sigma^2/\ev{N}$ at large $\ev{N}$ and small $W<4$ is due to the size effect: the radius of our disk becomes larger than the maximum imaginary part of eigenvalues.

Let us return to the linear dependence $N_c(L,W) \propto L$ found above. 
Similar dependence is known for the metallic phase of the Anderson model. 
Indeed, the conductance $G(L)$ of a metallic sample is determined by the number of energy levels $N_c$ within the interval $E_c$. 
On the other hand, a metallic cube with edge $L$ has conductance $G(L)$ proportional to $L$. 
Thus, for the Anderson model $N_c \propto L$ as well.

This analogy may be used to speculate about the total transmission of the random laser cube $g(L)$~\cite{fisher:1981,feng:1988} described by TME model in the delocalized diffusion regime $W_c L^{-1/2} < W < W_c$ .
One may speculate that $g(L)$ is proportional to $N_c(L,W)$ and the mean square mesoscopic fluctuations of $g(L)$ from one random realization of the random laser media cube to another $\delta g(L)$ is determined by the corresponding Ginibre complex ensemble value $[N_c(L,W)/\pi]^{1/2} \propto L^{1/2}$, i. e.  
\begin{equation}\label{eq:dg} 
\delta g(L)/ g \sim L^{-1/2},
\end{equation}
much larger than $1/L$ for universal mesoscopic fluctuations in Hermitian case. One can argue for such dependence on $L$ also in the following way. Delocalized wave functions average out effect of $\text{Im} E_i$, such that the average for a given realization is $\ev{\text{Im} E_i} \sim W/L^{3/2}$. Such a small positive (negative) $\ev{\text{Im} E_i}$ leads to amplification (absorption) by the cube of the order $\exp[\ev{\text{Im} E_i} L]$. 
At large $L$ we get $\ev{\text{Im} E_i} L \propto 1/L^{1/2}$ leading to Eq.~\eqref{eq:dg}.

\section{Acknowledgement}
We are grateful to A. Kamenev and I. K. Zharekeshev for useful discussions. Calculations by Y.H. were supported primarily by the National Science Foundation through the University of Minnesota MRSEC under Award Number No. DMR-1420013 and DMR-2011401. The authors acknowledge the Minnesota Supercomputer Institute (MSI) at the University of Minnesota for providing resources that contributed to the research results reported within this paper.
\medskip


%

\end{document}